\documentclass[twocolumn,showpacs,floats,floatfix,superscriptaddress,aps,pra]{revtex4-1}
\usepackage{amssymb}
\usepackage{mathrsfs}
\usepackage{graphicx}
\usepackage{float}
\usepackage[caption=false]{subfig}
\usepackage[pdftex,colorlinks=red]{hyperref}

\usepackage{epstopdf}

\usepackage[normalem]{ulem}
\usepackage{verbatim}
\usepackage{xcolor}
\usepackage{bm}
\usepackage{amsmath, amssymb}
\usepackage{makecell}
\usepackage{multirow}

\begin{document}

\title{Filling up complex spectral regions through non-Hermitian disordered chains}

\author{Hui Jiang}  \email{phyhuij@nus.edu.sg}
\affiliation{Department of Physics, National University of Singapore, Singapore 117551, Republic of Singapore}

\author{Ching Hua Lee}  \email{phylch@nus.edu.sg}
\affiliation{Department of Physics, National University of Singapore, Singapore 117551, Republic of Singapore}

\begin{abstract}
Eigenspectra that fill regions in the complex plane have been intriguing to many, inspiring research from random matrix theory to esoteric semi-infinite bounded non-Hermitian lattices. In this work, we propose a simple and robust ansatz for constructing models whose eigenspectra fill up generic prescribed regions. Our approach utilizes specially designed non-Hermitian random couplings that allow the co-existence of eigenstates with a continuum of localization lengths, mathematically emulating the effects of semi-infinite boundaries. While some of these couplings are necessarily long-ranged, they are still far more local than what is possible with known random matrix ensembles. Our ansatz can be feasibly implemented in physical platforms such as classical and quantum circuits, and harbors very high tolerance to imperfections due to its stochastic nature.
\end{abstract}

%\begin{abstract}

%\end{abstract}
\maketitle

%\clearpage
%%%%%%%%%%%%%%%%%%%%%%%%%%%%%%%%%%%%%%%%%%%%%%%%%%%%%%%%%%%%%%%%%%%%
%\onecolumngrid
%\newpage

%\renewcommand{\thefigure}{S\arabic{figure}}
%\renewcommand{\thetable}{S\arabic{table}}
%\renewcommand{\theequation}{S\arabic{equation}}
%\renewcommand{\thefigure}{S\arabic{figure}}
%\renewcommand{\thetable}{S\arabic{table}}
%\renewcommand{\thefigure}{S\arabic{figure}}
%\renewcommand{\thetable}{S\arabic{table}}
\setcounter{equation}{0}
\setcounter{figure}{0}
\setcounter{table}{0}
%%%%%%%%%%%%%%%%%%%%%%%%%%%%%%%%%%%%%%%%%%%%%%%%%%%%%%%%%%%%%%%%%%%%

%\begin{center}
  %  {\bf \large {SM for filling PBC loop....}}
%\end{center}
%\tableofcontents
% \tableofcontents
%%%%%%%%%%%%%%%%%%%%%%%%%%%%%%%%%%%%%%%%%%%%%%%%%%%%%%%%%%%%%%%%%%%%
\section{ Introduction}
The spectra of non-Hermitian systems lie in the 2D complex plane, and can exhibit intriguing geometric and topological spectral transitions~\cite{berry2004EP,PhysRevLett.80.5243,experimentzeuner2015,experimentHodaei2017,experimentChen2017,experimentzhu2018,experimentgal2018,experimentmiguel2018,experimentli2019,experimentwu2019,lee2020exceptional,li2021quantized,circuitPhysRevLett.126.215302,PhysRevX.8.031079,IPRPhysRevA.95.022117,RevModPhys.93.015005}. 
In particular, it is known~ \cite{PhysRevLett.124.086801,PhysRevLett.124.056802,PhysRevLett.125.126402,PhysRevB.104.125109}
 that if the spectrum under periodic boundary conditions (PBCs) is a loop that encloses a nonvanishing region, the spectrum of the same system under semi-infinite boundary conditions (SIBCs) will fill up the interior of this loop.
 This intriguing fact is due to the non-local nature of the non-Hermitian skin effect (NHSE), which has inspired numerous theoretical and experimental~\cite{circuitliu2021non,experimentsebastian2020,xiao2020non, experimentlin2021steering,circuithofmann2020reciprocal,experimenthelbig2020,circuitPhysRevLett.122.247702,experimentBudich2020,PhysRevResearch.2.013058,circuithofmann2020reciprocal,PhysRevApplied.14.064076,PhysRevResearch.2.013280,circuitrafi2021non,experimentMcDonald2020,circuithofmann2020reciprocal,circuitzou2021observation,zhang2021observation} developments and challenged various longheld paradigms in physics.
The NHSE, which arises in non-Hermitian lattices with broken reciprocity, amplifies and pumps all states towards a boundary, such that the effects of boundary hoppings become non-perturbatively large~\cite{PhysRevLett.121.086803,PhysRevResearch.1.023013,PhysRevLett.124.066602,PhysRevLett.125.118001,PhysRevB.102.085151,PhysRevLett.121.086803,PhysRevResearch.1.023013,Longhi:20,li2020critical,PhysRevLett.125.118001,PhysRevLett.124.066602,PhysRevB.104.L161106,PhysRevB.99.201103,circuitMotohiko20191,PhysRevLett.121.086803,PhysRevB.102.205118,PhysRevLett.125.118001,PhysRevB.102.241202,PhysRevLett.124.250402,PhysRevB.102.201103,PhysRevB.104.L161106,lee2021many,PhysRevB.103.075126,PhysRevB.104.125416,PhysRevLett.126.176601,li2021impurity,PhysRevLett.127.116801,circuitxuke2021,PhysRevB.104.L161106,PhysRevB.103.085428,PhysRevB.104.165117,zhang2021universal,yoshida2019mirror,PhysRevLett.124.250402,Lee2019hybrid,zhang2020non,PhysRevB.102.085151,luo2020skin,YUCE2020126094,PhysRevB.99.245116,PhysRevB.100.075403,luo2020skin,PhysRevLett.125.186802}.
  
Intuitively, a semi-infinite 1D lattice system can have a spectrum that fills up a 2D region because its eigenstates only need to satisfy boundaries conditions on one side, and are hence free to accumulate against it with any spatial decay length. As such, an eigenstate is characterized by two continuous variables: its wavenumber and decay length, the latter which possess no Hermitian analog.
 However, true semi-infinite systems can neither be numerically nor experimentally simulated, and their non-Hermitian properties have so far been mathematical curiosities.

In this work, we show how to construct finite 1D systems whose spectra nevertheless fills up the 2D interiors of their PBC spectra. This is achieved with appropriately designed disordered couplings which mathematically simulate the effects of semi-infinite boundaries, namely the co-existence of a continuum of different decay length scales of the skin eigenstates. Notably, the density of states in the 2D complex plane can be fine-tuned towards a variety of desired profiles by tuning the disorder distribution. While it is arguably easy to fill up a 2D spectral region with the eigenenergies of many separate (uncoupled) 1D chains, doing so with a \emph{single} chain is nontrivial due to the non-local effects of NHSE accumulation that can propagate across very distant parts of the chain~\cite{li2021impurity,PhysRevLett.127.116801}. As such, our construction can be construed as a stochastic means to subtly control the distribution of skin decay lengths, and also the propagation of skin accumulation tendencies.  

%\section{2D SPECTRAL DISTRIBUTION FROM A 1D NONHERMITIAN LATTICE}
\section{Exploring the complex energy plane by tuning boundary conditions}

The starting point of our work is the observation that, by modifying the boundary conditions of a non-Hermitian system with unbalanced hoppings, we can access a continuum of complex energy spectra, and thus sweep across the interior of the PBC loop~\cite{PhysRevLett.126.176601,PhysRevB.103.L140201,li2021impurity,PhysRevLett.127.116801}. This is thanks to the non-Hermitian lattice's extreme sensitivity to the boundary conditions, a phenomenon commonly known as the non-Hermitian skin effect. In the extreme limit of open boundary conditions (OBCs), the eigenenergy continuum are given by
\begin{equation}\label{eq1-1}
\begin{split}
\bar E(k)=E(k+i\kappa_{\text{c}})\ ,
\end{split}
\end{equation}
where $E(k)$ is the dispersion of the original Hamiltonian $H(k)$, and $\kappa_{\text{c}}$ is the imaginary part of the momentum that represents the boundary localization of former bulk eigenstates~ \cite{PhysRevB.104.125109,circuitliu2021non,experimentsebastian2020,xiao2020non,PhysRevResearch.1.023013,PhysRevLett.124.066602,PhysRevLett.125.118001,PhysRevB.102.085151,PhysRevLett.123.170401,PhysRevLett.123.066404}
, which now decay like $\sim e^{-\kappa_\text{c}x}$. In general, $\kappa_{\text{c}}$ can be a complicated function of $k$, and is determined by the condition that $\bar E(k)$, $k\in [0,2\pi)$ does not enclose any nonzero area~\cite{grein1971multiconfiguration,PhysRevLett.121.086803,PhysRevLett.123.066404,PhysRevLett.123.246801,PhysRevResearch.1.023013,xiao2020non,PhysRevLett.124.066602,PhysRevLett.125.226402,wu2021connections,guo2021non,experimentlin2021steering}. But for the purpose of this work, $\kappa_\text{c}$ remains as a constant.
 %And while the system $H_{\kappa_{\text{c}}}$ with $\kappa_{\text{c}}\neq 0$, there must have PBC loop,  which represents the winding of “energy” (eigenvalue of Hamiltonian) on the complex plane as momentum traverses the Brillouin zone ($\kappa_{\text{c}}=0$) with periodic boundary condition, and no winding of “energy” as momentum traverses the general Brillouin zone with $H_{\kappa_{\text{c}}}$ in non-Hermitian systems.
 
  Interpolating $\kappa$ from $\kappa_{\text{c}}$ to $0$, the energy spectrum $E(k+i\kappa)$ will start off as the OBC energy spectrum, and continuous evolve till it coincides with the PBC spectrum at $\kappa=0$, in the process passing through the entire region enclosed by the PBC loop. Below, we describe how to physically access such intermediate values of $\kappa$ by tuning the boundary hoppings.  
	%kappa is not called the minimal coupling - usually that is only for real gauge fields
	
  Using the Hatano-Nelson model~\cite{PhysRevX.8.031079,PhysRevB.104.L161106,NHMPhysRevLett.77.570,NHMhatano1997vortex,PhysRevB.92.094204,PhysRevB.103.L140201} for convenient illustration, since it only has nearest neighbor unbalancing hoppings, we have
  \begin{equation}\label{eq1-2}
\begin{split}
H=\sum^{L-1}_{n=1} t |n\rangle\langle n+1|+   t'|n+1\rangle\langle n|& \\
+\mu |L \rangle\langle 1|+ \mu' |1\rangle\langle L|&\  ,
\end{split}
\end{equation} 
 where $t/ t'$ is the left/right hopping amplitude and $\mu,\mu'$ represent the boundary hoppings that connect the left and right boundaries. 
Writing an eigenstate as $|\psi\rangle=\sum_n\psi_n|n\rangle$ with $H|\psi\rangle=E|\psi\rangle$, the wave function amplitude is given by $\psi_n $ at the $n$-th site in the lattice.
For a lattice comprising $L$ sites ($n = 1, 2, 3, ..., L $), the matrix Hamiltonian $H$ in Eq.\eqref{eq1-2} gives rise to bulk and boundary conditions which together give (See Appendix \ref{apppendix1}):
\begin{equation}\label{eq1-3}
\begin{split}
&\left(tt'\left(\frac{t'}{t}\right)^{L}\frac{1}{z^L}- t' \mu\right)\left(t\mu' z^{L}-tt'\right) \\
&\ =z^2\left(t\mu' \left(\frac{t'}{t}\right)^{L}\frac{1}{z^L}-tt'  \right)\left(t^2z^L-t \mu \right)\ ,
\end{split}
\end{equation} 
%\begin{equation}\label{eq1-3}
%\begin{split}
%\left(tt'\left(\frac{t'}{t}\right)^{L}\text{e}^{\kappa L- ikL}- t' \mu\right)\left(t\mu' \text{e}^{-\kappa L+ikL}-tt'\right)& \\
%=\text{e}^{-2 \kappa+2ik}\left(t\mu' \left(\frac{t'}{t}\right)^{L}\text{e}^{\kappa L- ikL}-tt'  \right)\left(t^2\text{e}^{-\kappa L+ikL}-t \mu \right)&\ ,
%\end{split}
%\end{equation} 
with $E=tz+t'/z$ and $z=\exp(-\kappa+ik)$. Solving this characteristic equation gives $\kappa$, as shown in TABLE.\ref{tab1}, which can then be substituted into $\bar E=E(k+i\kappa)$ to yield the spectrum (See Appendix \ref{apppendix1}). It is seen that under OBCs, $\kappa$ only depends on the relative amplitudes $t/t'$ of the left/right hoppings. However, when the boundary and bulk hoppings are not equal, $\kappa$ will depend either on $t/\mu$ or $t'/\mu'$, depending on the relative strength of $t$ and $t'$ as well as whether $\mu\mu'$ or $tt'$ is larger.

\begin{table}[h!]
\renewcommand\arraystretch{1.5}
  \centering
  \begin{tabular}{c|c|c}
\hline
\hline
 &$t'<t$& $t'>t$\\
 \hline
\multirow{2}{*}{$\mu\mu' < tt'$}  &$ \kappa=\log\left({t/\mu}\right)/L$
&$ \kappa=\log\left({\mu'/t'}\right)/L$\\   
&$(\kappa\in[0,\kappa_{\text{c}}])$
&$(\kappa\in[\kappa_{\text{c}},0])$\\
\hline
$\mu\mu' > tt'$ &$ \kappa=\log\left({\mu'/t'}\right)/L$&$\kappa=\log\left({{t/\mu}}\right)/L$\\
\hline
 \multicolumn{3}{l}{PBC \ ($\mu=t,\mu'=t'$): \qquad\  $\kappa=0$}\\
 \multicolumn{3}{l}{OBC\ ($\mu=\mu'=0$):\qquad\  $\kappa=\kappa_c=1/2\log(t/t')$,}\\
\hline
\hline
\end{tabular}
\caption{  \label{tab1} Approximate analytic forms of $\kappa$ of the model Eq.\eqref{eq1-1} in various regimes, for large $L$. Here, $\kappa_\text{c}=1/2\log(t/t')$, the value of $\kappa$ where the spectrum collapses into the OBC spectrum that encloses zero area. With appropriate tuning of boundary hoppings $\mu,\mu'$, we will be able to obtain any $\kappa$ that lies between the PBC and OBC cases i.e. $[0,\kappa_\text{c}]$.\\
%\CH{CH: Is there a need for the lim sign? If there is a lim, $L$ should not be present at the RHS, which should just be zero.}
%\JH{JH: the calculation of  $\kappa $  must use approximate conditions $\lim\limits_{L\gg 1} x^{1/L}\to 1$ with $x>1$.Would it be better If I change $\lim\limits_{L\to \infty} $ to $\lim\limits_{L\gg 1} $?}}
%: CH: I have simply mentioned that expressions are for large L. Not strictly correct to use limits, since we have an asymptotic expansion.
}
\end{table}

\begin{figure}[htp]
    \begin{centering}
    \includegraphics[width=1\linewidth]{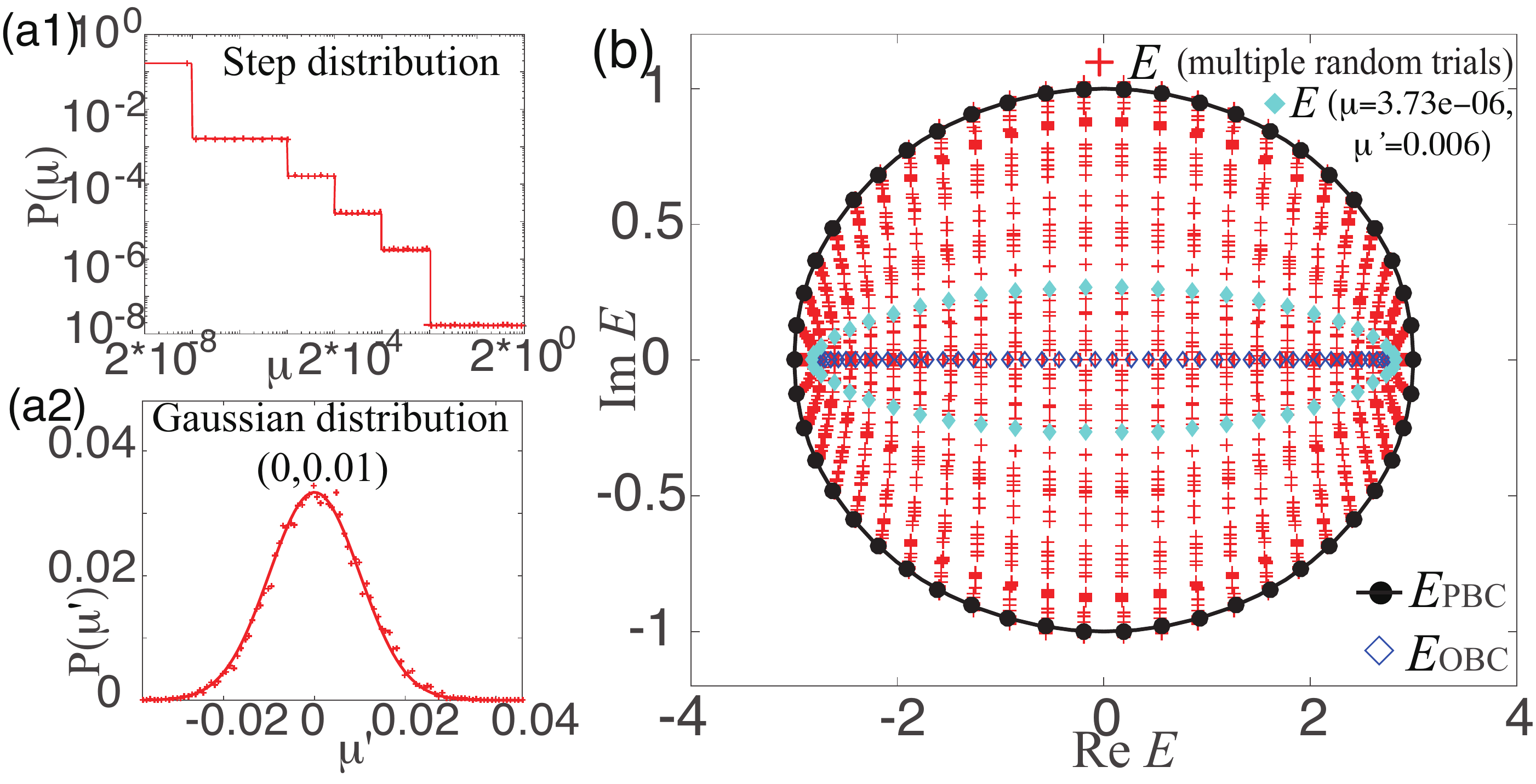}
    \par\end{centering}
    \protect\caption{\label{fig:1} 
\textbf{Filling up a spectral ellipse with a large number of separate Hatano-Nelson chains with random boundary hoppings.} (a1-a2) shows the random distributions of $\mu,\mu'$ boundary hoppings, while (b) shows the combined energy spectra $E$ (red crosses) from 40 different random trials, which fills up the interior of the elliptical PBC loop of the model given by Eq.\eqref{eq1-2} with parameters $t=2,t'=1,L=50$. From Table I, only the distribution $P(\mu)$ affects the filling, which here forms ``bands'' of approximately equal density due to the chosen step-like distribution of $P(\mu)$. The energies from an illustrative $(\mu,\mu')$ pair is indicated in light blue.
%	$$P(\mu')=\left\{
%	\begin{aligned}
%	&10^{-7}/6\qquad (\mu'\in(10^{-2},1])\\
%&10^{-5}/6\qquad(\mu'\in(10^{-3},10^{-2}])\\
%&10^{-4}/6\qquad(\mu'\in(10^{-4},10^{-3}]),\\
%&10^{-3}/6\qquad(\mu'\in(10^{-4},10^{-5}])\\
%&10^{-2}/6\qquad(\mu'\in(10^{-7},10^{-5}])\\
%&1/6\qquad\qquad(\mu'\in(0,10^{-7}])
%	\end{aligned}
%	\right.
%	$$
}
\end{figure}

 The upshot is that if we take the boundary hoppings $\mu,\mu'$ to be random numbers from $(0,t)$ and $(0,t')$, 
 the energy spectrum will fall within the PBC spectral loop and, after multiple random trials, the combined energy spectra will fill up the loop, as shown in Fig.\ref{fig:1}(b). So far, this is not very surprising, since we are using an ensemble of 1D systems to fill up a 2D region. In the following sessions, we shall demonstrate how we can instead construct a \emph{single} 1D system whose eigenenergies fill up the interior of a spectral loop.
%Hereupon, adjusting the connection condition of the boundary can fill the PBC loop with multiple random trials.
%The 1D model with 2D spectral distribution, which fill the PBC loop with 1 random trial rather than  multiple times, will be discussed in the next section.

\section{Filling 2D spectral region with a 1D non-Hermitian lattice}

Previously, we filled the 2D spectral region enclosed by the PBC loop via an ensemble of 1D chains. However, to do the same with a \emph{single} 1D chain is a nontrivial feat, even for one constructed by concatenating many 1D chains. This is because the NHSE relentlessly pump all states towards one direction, including across the concatenated chains, thereby fundamentally modifying their individual nature. Specifically, we expect different behavior from boundary hoppings that close up the individual 1D chains, compared to those that connect many 1D chains into one long periodic chain. 

\begin{figure}[bp]
    \begin{centering}
    \includegraphics[width=1\linewidth]{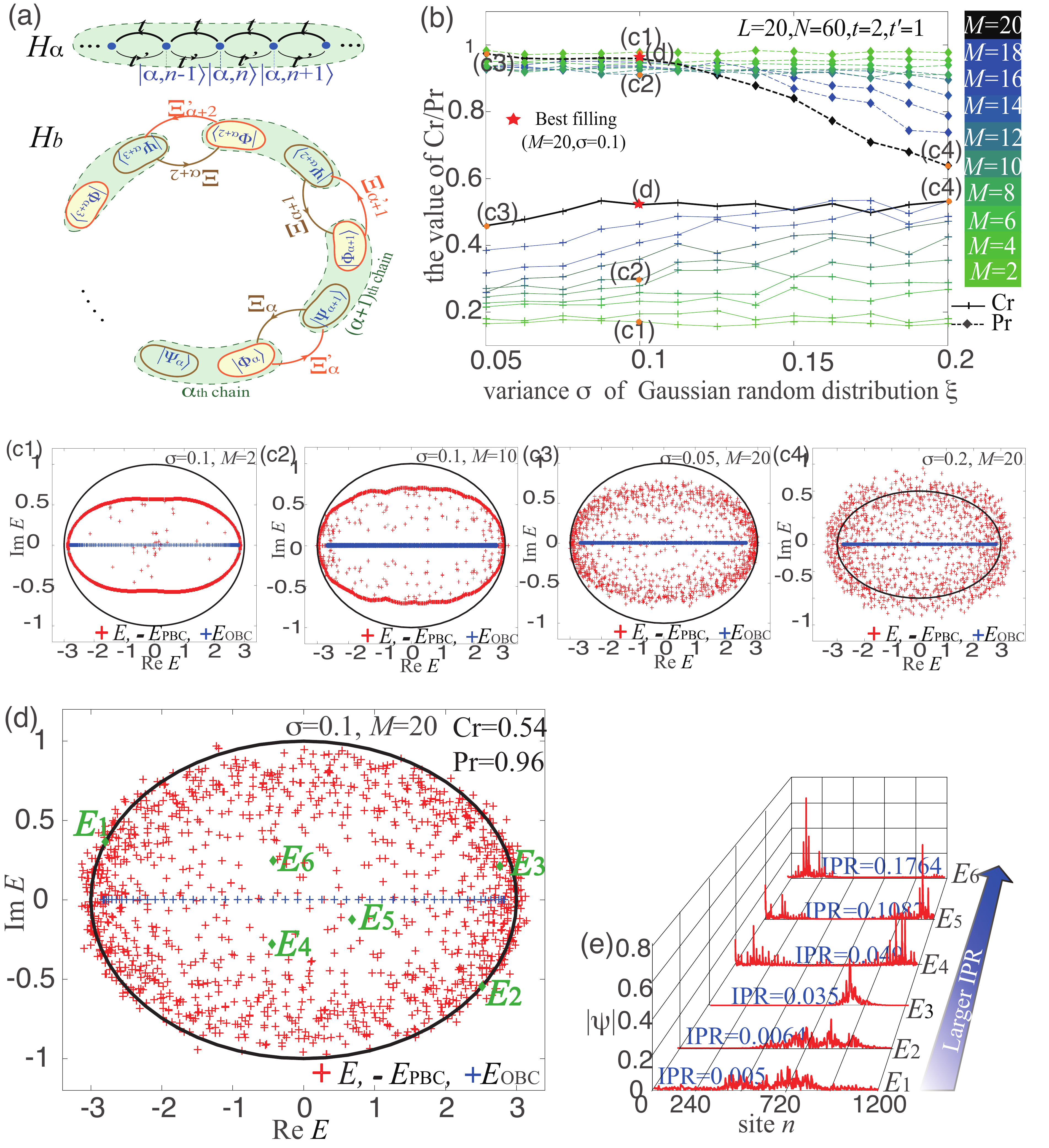}
    \par\end{centering}
    \protect\caption{\label{fig:2}\textbf{Filling of spectral region from a single long chain with disordered couplings.} 
		(a) Structure of our 1D chain model Eqs.(\ref{eq2-1}-\ref{eq2-3}), which consists of $N$ Hatano-Nelson chain segments (green and yellow) that are randomly coupled to adjacent chains via their first and last $M$ sites, as given by matrices $\Xi$ and $\Xi'$. The hopping amplitudes $\xi\in (0, \sigma)$ are Gaussian random distributed, and phases $\phi$ are uniformly distributed. (b) The dependence of the Coverage rate (Cr) Eq.\eqref{eq2-5} and Participation rate  (Pr)  Eq.\eqref{eq2-6} on the hopping amplitude variance $\sigma$, for different $M$. While Cr is generally insensitive to $\sigma$, it improves significantly with $M$. Pr remains almost complete at $1$ for $\sigma\leq 0.1$, beyond which it decreases. (c1-c4) shows the energy spectrum at different $\sigma$ and $M$ combinations as indicated in (b); as $M$ increases, interior eigenvalues within the spectral loop proliferate, finally resulting in a filled spectral interior. Excessive $\sigma$, however, causes the filling to exceed the PBC loop boundaries. (d) The best filling spectrum from (b), with representative eigenstates shown in (e). Generally, more localized states occur deeper in the loop interior. $E_{\text{OBC}}$ and $E_{\text{PBC}}$ refer to the eigenenergies of base model  $H_{\alpha}$ Eq.(\ref{eq2-2}). 
}
\end{figure}

To construct a bona fide 1D lattice whose spectrum does fill up a 2D spectral region, we instead consider a model $H$ of the form
\begin{equation}\label{eq2-1}
\begin{split}
H=&\sum^N_{\alpha=1}H_{\alpha}+H_b\ ,\\
\end{split}
\end{equation} 
where the first term is that of $N$ Hatano-Nelson chains of length $L$ without boundary hoppings i.e.   
\begin{equation}\label{eq2-2}
\begin{split}
&H_{\alpha}=\sum^{L-1}_{n=1} t |\alpha,n\rangle\langle \alpha, n+1|+   t'|\alpha,n+1\rangle\langle \alpha, n|\ ,
\end{split}
\end{equation} 
%\JH{JH: the model $H_{\alpha}$ is the length $L$ of OBC chain, with the end hopping $|\alpha,1\rangle\langle \alpha, 2|$ with $n=1$, and $ |\alpha,L-1\rangle\langle \alpha, L|$ with $n=L-1$. }
where $\alpha$ is the chain index,  and $n$ labels the sites in the $\alpha$-th chain. %The $N$ successive chains are PBC-linked to form a long 1D system with $NL$ sites (FIG.~\ref{fig:2}a): $|\alpha+N,n\rangle=|\alpha,n\rangle$ ($n=1,2,..,L$).} 
The second term $H_b$ contains all the random couplings between adjacent chains $\alpha$ and $\alpha+1$, and takes the form
 \begin{equation}\label{eq2-3}
H_b=\sum_{\alpha=1}^{N}|\Phi_{\alpha}\rangle\Xi_{\alpha} \langle\Psi_{\alpha+1}|+   |\Psi_{\alpha+1}\rangle\Xi'_{\alpha}\langle \Phi_{\alpha} |  \ ,\\
\end{equation} 
where  $|\Phi_{\alpha}\rangle$ and $|\Psi_{\alpha}\rangle$ respectively contains the first $M$ and last $M$ sites of the $\alpha$-th chain i.e.
\begin{equation}\label{eq2-4}\begin{split}
 |\Phi_{\alpha}\rangle&=\left( |\alpha,1\rangle,|\alpha,2\rangle,...,|\alpha,M\rangle\right)\ ,\\
|\Psi_{\alpha}\rangle&=\left( |\alpha,L-M+1\rangle,...,|\alpha,L-1\rangle,|\alpha,L\rangle\right)\ ,\\
\end{split}\end{equation} 
and $\Xi_{\alpha},\Xi'_{\alpha}$ are $M\times M$ random matrices whose elements represent their random couplings between the chains. Together, they connect the chains into a long PBC loop via $2NM^2$ random couplings. As elaborated later, the coupling length $M\in [1,L]$ profoundly affects the filling, as demonstrated in Fig.\ref{fig:2}(c) and subsequent figures. 

We choose the random couplings matrix elements $\xi\text{e}^{i\phi}$ of $\Xi_{\alpha},\Xi'_{\alpha}$ from a random ensemble with amplitudes $\xi$ Gaussian distributed with mean $0$ and variance $\sigma$, and phase $\phi$ uniformly distributed [Fig.\ref{fig:2}(a)]. Since the spectrum changes dramatically when inter-chain couplings are small, analogous to the boundary couplings discussed in the previous section, the coupling amplitude variance $\sigma$ significantly affects the filling behavior [Fig.\ref{fig:2}(b)].

%CH: restructed the following paragraphs, so that it doesnt sound like Cr and Pr are only suitable for characterizing M and disorder distributions only.
To quantify how completely and evenly the spectral loop region is filled by the eigenenergies, we introduce two metrics: Cr, the coverage rate and Pr, the participation rate. % The coupling matrix size $M$ and random distribution form $\xi\text{e}^{i\phi}$ of  matrix $\Xi_{\alpha},\Xi'_{\alpha}$ can affect the energy spectrum which filled up a solid region within part or the whole PBC loop. Therefore, how to quantify the energy spectrum distribution of  disordered system is particularly important. In addition to visually observing the energy spectrum filling situation(FIG.\ref{fig:2}(c1-c3)), the larger the $M$, the better the energy spectrum filling, we can also use physical quantities to describe it. \\
%The first is the impact on the coupling matrix size $M$, 
To define the coverage rate (Cr), we divide the interior region of the PBC loop (spectrum of $H_{\alpha}$ under PBCs) into $\mathcal{N} $ parts (for a large $\mathcal{N} $); then we count the number $\mathcal{N}'$ of parts which contain one or more eigenvalues of $H$ Eq.\eqref{eq2-1} within itself. The ratio
\begin{equation}\label{eq2-5}
\begin{split}
\text{Cr}=\frac{\mathcal{N}'}{\mathcal{N}} \ ,
\end{split}
\end{equation} 
is the coverage rate Cr. The larger the Cr, the more complete is the energy filling; a small Cr indicates that the filling occurs very inhomogeneously. This definition of Cr remains meaningful even when the PBC loop is irregular and it is hard to directly see how well it is filled by the eigenenergies. Next, we also define the participation rate (Pr), which represents the fraction of eigenenergies of $H$ that are within the PBC loop of $H_\alpha$ i.e.
\begin{equation}\label{eq2-6}
\begin{split}
\text{Pr}=\frac{\text{num}(E)}{NL}\ ,
\end{split}
\end{equation} 
where $\text{num}(E) $ is the number of eigenenergies within the PBC loop, and $NL$ is the total number of sites in our lattice model Eq.\eqref{eq2-1}. A high participation rate indicates that few eigenenergies are outside the PBC loop.

From Fig.\ref{fig:2}(b), it is evident that the longer the range $M$ of the inter-chain random couplings, the better the coverage Cr. This is because short-ranged inter-chain couplings in $\Xi_{\alpha},\Xi'_{\alpha}$ only couple sites close to the end of the chains, and the entire lattice is still akin to a long PBC chain with somewhat complicated couplings. Indeed, at small $M$, the spectrum of $H$ is still a well-defined loop [Fig.\ref{fig:2}(c1)], which by definition cover the PBC interior region very poorly. As $M$ increases, more distant sites between adjacent chains are randomly coupled, and some eigenenergies start to appear in the interior of the spectral loop of $H$ [Fig.\ref{fig:2}(c2)]. They almost always appear in its interior because disorder generically break translation invariance, thereby allowing for localized skin mode accumulation. These localized modes have larger effective $\kappa$ by virtue of their shorter decay lengths, and hence tend towards the interior of the spectral loop. This is further examined in the next section; here we mention that intuitively, we expect these random coupling-induced localized skin modes energies to be closer to the real line because the net effect of many random couplings is to prevent any state from being amplified or attenuated too many times consecutively. Finally, the coverage Cr reaches its maximum [Fig.\ref{fig:2}(b)] when $M=L$ i.e. when all sites in our lattice Eqs.(\ref{eq2-1}-\ref{eq2-3}) are randomly coupled. In this limit, all vestiges of a PBC spectral loop are gone, and we observe a continuous eigenstate density within the PBC spectral loop of $H_\alpha$. 

The filling of the 2D spectral region can be further optimized by tweaking the probability distribution of the individual inter-chain hopping strengths $\xi e^{i\phi}$. We shall keep the phase $\phi$ as being uniformly distributed, and just vary the variance $\sigma$ of the Gaussian-distributed amplitude $\xi$ (with zero mean). As shown in Fig.\ref{fig:2}(c), the cloud of eigenenergies become larger as $\sigma$ grows, since stronger random couplings invariably perturb their ``trapped'' localized skin modes energies more strongly. We obtain the best filling with $M=L$, and when the eigenenergy cloud just fills the PBC spectral loop of $H_\alpha$ without going out of it. This occur at a critical value of $\sigma=\sigma_{\text{c}}$, where the participation ratio Pr just starts to decrease from 1 [Fig.\ref{fig:2}(b)].

%Then considering the impact on the random distribution $\xi\text{e}^{i\phi}$ from matrix $\Xi_{\alpha},\Xi'_{\alpha}$,  only Cr cannot describe filling, we redefine other quantities--The inflection point of the Participation rate (Pr) curve corresponds to the best filling $(0,\sigma_{\text{c}})$ within  the whole PBC loop, in FIG.\ref{fig:22}(c,c1-c3).\\

  \begin{figure}[t!]
    \begin{centering}
    \includegraphics[width=1\linewidth]{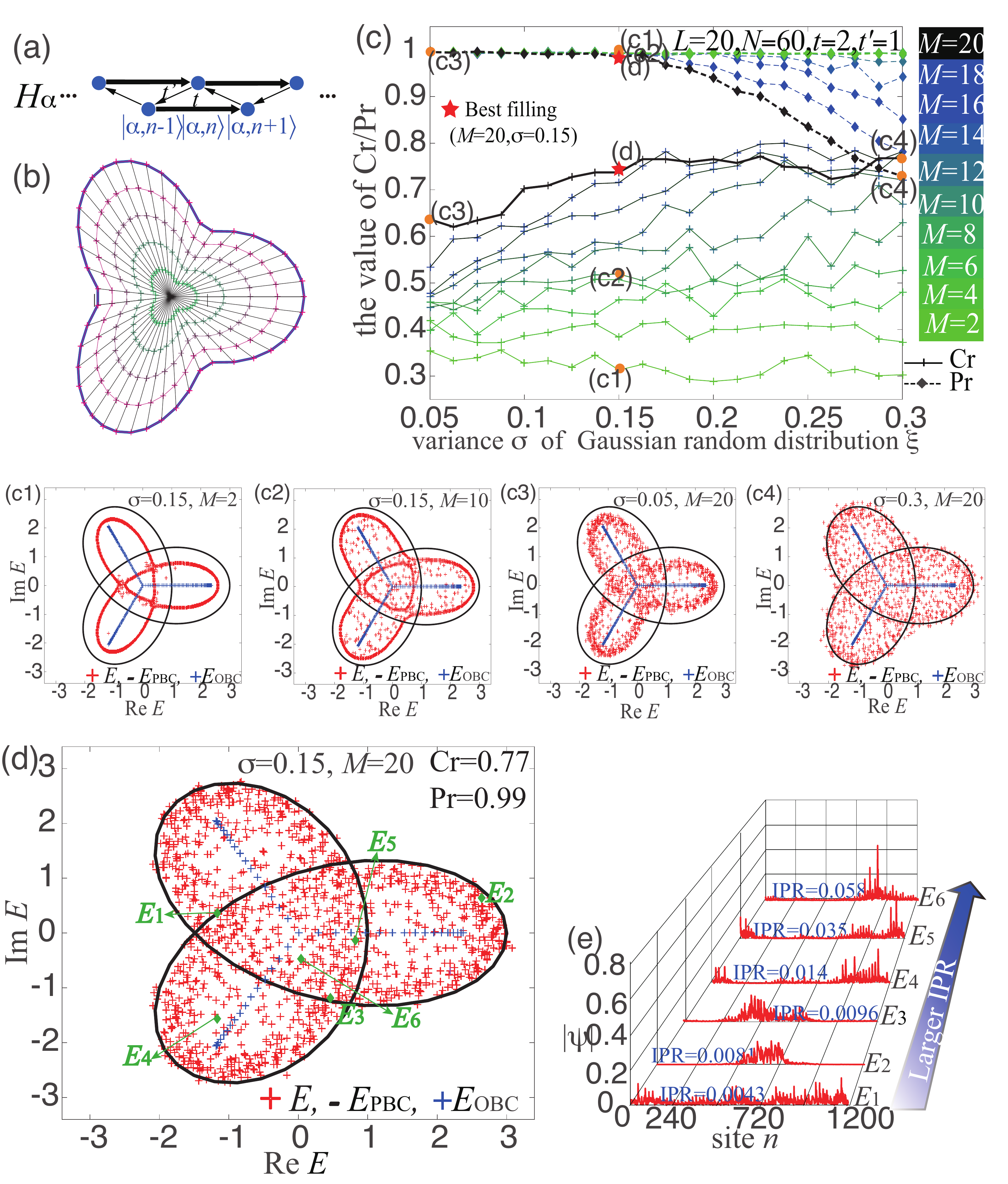}
    \par\end{centering}
    \protect\caption{\label{fig:3}
\textbf{Filling of spectral region from a single long heterogeneous chain with disordered couplings.} (a) Base lattice structure of our model Eq.(\ref{eq3-3}) with next-nearest neighbor hoppings; random hoppings linking them (shown in Fig.\ref{fig:2}) having amplitudes $\xi\in (0, \sigma)$ that are Gaussian random distributed, and phases $\phi$ that are uniformly distributed. (b) Division of the irregularly shaped PBC loop into  large $\mathcal{N}$ parts for the computation of Cr, the Coverage rate. (c) Dependence of the Coverage rate (Cr) Eq.\eqref{eq2-5} and Participation rate  (Pr)  Eq.\eqref{eq2-6} on the hopping amplitude variance $\sigma$, for different $M$. While Cr increases only slightly with $\sigma$, it improves significantly with $M$. Pr remains almost complete at $1$ for $\sigma\leq 0.15$, beyond which it decreases. (c1-c4) shows the spectrum at different $\sigma$ and $M$ combinations as indicated in (c); as $M$ or $\sigma$ increase, interior eigenvalues within the spectral loop proliferate, finally resulting in a filled spectral interior. (d) The best filling spectrum from (c), with representative eigenstates shown in (e). Generally, more localized states occur deeper in the loop interior. $E_{\text{OBC}}$ and $E_{\text{PBC}}$ refer to the eigenenergies of base model  $H_{\alpha}$ Eq.(\ref{eq3-3}). 
}
\end{figure}

So far, we have only used the Hatano Nelson model Eq.(\ref{eq2-2}) as the base model $H_{\alpha}$. As a model with a simple elliptical PBC loop, it is an appropriate paradigmatic model for disorder spectral filling. However, our approach also works for generic models with nontrivial spectral winding loops.  

For instance, let's add to the disorder coupling $H_b$ Eq.\eqref{eq2-3} a different base Hamiltonian given by~\cite{PhysRevB.102.085151}
 \begin{equation}\label{eq3-3}
\begin{split}
&H_{\alpha}=\sum^{L-1}_{n=1} t |\alpha,n\rangle\langle \alpha, n+a|+   t'|\alpha,n+b\rangle\langle \alpha, n|\ ,
\end{split}
\end{equation} 
which has hoppings $a$ sites to the left with amplitude $t$, and hoppings $b$ sites to the right with amplitude $t'$, as illustrated in Fig.\ref{fig:3}(a). These $H_\alpha$ form $N$ identical chains with open boundary conditions, connected to each other by disordered couplings just as before Eqs.(\ref{eq2-1},\ref{eq2-3}). 

In the same way, we can also get the best filling conditions. The larger the Cr, the better the spectral region filling, in Fig.\ref{fig:3}(c). There have same results with the previous Hatano-Nelson model,  the value of $M$ is closer to $L$, the better PBC loop-filled energy spectrum. %And the inflection point of the Participation rate (Pr) curve (FIG.\ref{fig:3}) corresponds to the best filling $(0,\sigma_{\text{c}})$ within  the whole PBC loop, in FIG.\ref{fig:3}(d).
Similarly,  the model with the best loop-filled energy spectrum can be obtained ($M=L$, and the optimal distribution of random couplings $\xi\exp(i\phi)$ for  Fig.\ref{fig:3}(d) can be obtained by considering the coverage rate (Cr) and participation rate (Pr). 

\section{Spectral filling and skin localization}
  \begin{figure}[htb]
    \begin{centering}
    \includegraphics[width=0.95\linewidth]{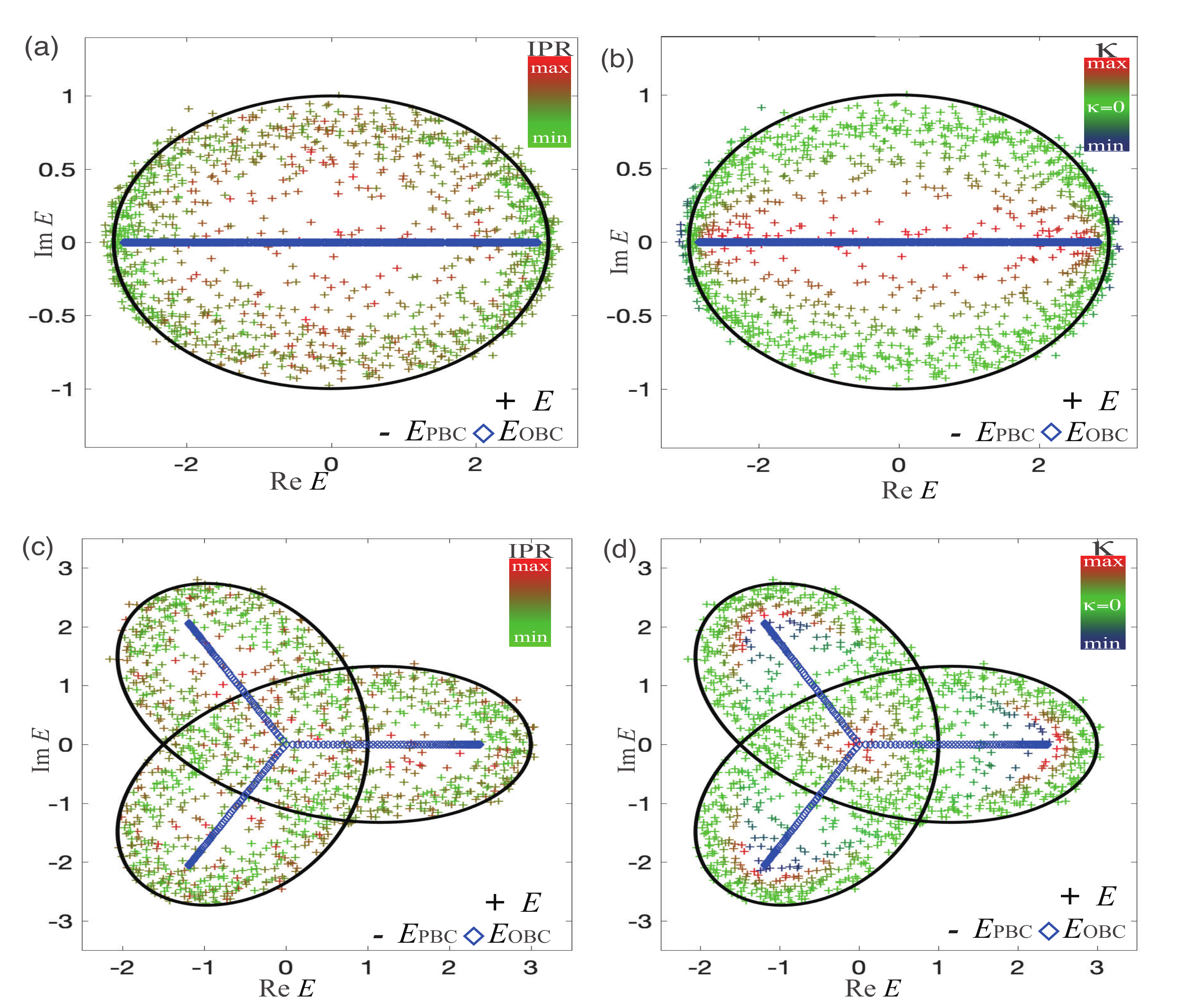}
    \par\end{centering}
    \protect\caption{\label{fig:4} \textbf{Locality (IPR) vs. inverse skin depth $\kappa$.} (a) IPR Eq.\eqref{eq3-1} and (b) $\kappa$ Eq.\eqref{eq3-2} of each eigenstate of our disordered chain Eqs.(\ref{eq2-1}-\ref{eq2-3}) at parameters of best filling ($\sigma=0.1,M=20$). Generally, small $\kappa$ (green), which correspond to delocalized Bloch states in the base model Eq.(\ref{eq2-2}), indeed correlate with relatively delocalized states in the disordered chain (low IPR, green).
(c) and (d) shows the same plots with the base Hamiltonian described by Eq.\eqref{eq3-3}, with similar conclusions.}
\end{figure}

A major inspiration for our approach has been the semi-infinite boundary condition, which is consistent with the coexistence of a continuum of eigenstate decay lengths (skin depths)~\cite{PhysRevLett.125.126402,PhysRevB.104.125109,circuitliu2021non,xiao2020non,PhysRevLett.121.086803,
PhysRevB.99.201103,
PhysRevResearch.1.023013,Longhi:20,li2020critical,PhysRevLett.125.118001,PhysRevLett.124.066602,PhysRevLett.123.066404,PhysRevB.100.035102}. Within our approach, the filling of the spectral loop is indeed intimately related to the realization of a continuum of spatial localization lengths. To quantify localization, we recall the definition of the inverse participation ratio (IPR)~\cite{IPRRevModPhys.80.1355,PhysRevB.103.014201,IPRPhysRevA.95.022117,PhysRevB.100.054301,PhysRevB.97.121401,PhysRevLett.122.237601}
\begin{equation}\label{eq3-1}
\begin{split}
\text{IPR}= \frac{\sum\limits_{\alpha,n}|\psi_{\alpha,n}|^4}{\left(\sum\limits_{\alpha,n}|\psi_{\alpha,n}|^2\right)^2}\ ,
\end{split}
\end{equation} 
with $|\Psi\rangle=\sum\limits_{\alpha,n}\psi_{\alpha,n}|\alpha,n \rangle$ a chosen eigenstate of the full Hamiltonian $H$. If a state is perfectly localized on only one site, the IPR takes the maximal value 1. In contrast, if a state is uniformly spread over $NL$ states,  $\text{IPR}=(NL)^{-1}\rightarrow 0$.

We next consider what can be reasonable expectation of the extent of localization for states within the spectral loop. States in a disordered system are invariably randomly shaped, but it is conceivable that they inherit the localization length of the clean background system $H_\alpha$, stochastically speaking. For a particular state with energy $E$, its expected (clean) skin depth $\kappa^{-1}$ is given through
\begin{equation}\label{eq3-2}
\begin{split}
E_{\alpha}(k+i\kappa)=E,
\end{split}
\end{equation}
where $k$ is an unimportant wavenumber and $E_{\alpha}$ is the energy of the 1-component model  $H_{\alpha}$. On the PBC loop, the clean system harbors Bloch states, and $\kappa=0$. Generally, states further from the PBC loop correspond to larger $\kappa$, and should be more localized.

In Fig.\ref{fig:4}, we indeed observe a correlation between weaker expected localization $\kappa$ and weaker numerically determined localization (smaller IPR) of the actual eigenstates. Conversely, for eigenenergies closer to $E_{\text{OBC}}$, where $\kappa\to \kappa_{\text{c}}$,  the eigenstates tend to be most localized (large IPR, red). This is also consistently observed for the base model $H_{\alpha}$ Eq.\eqref{eq3-3} with asymmetric hopping distances, albeit with slightly weaker correlation. \\

\section{Discussion}

In this work, we have devised a way to construct disordered 1D non-Hermitian chains  Eqs.(\ref{eq2-1}-\ref{eq2-3}) exhibiting eigenspectra that fill up the interiors of 2D regions in the complex energy plane. The filling extent and density can be adjusted by tuning the probability distributions of the random couplings, and effectively simulates the effects of SIBCs, which are physically unattainable.

It is interesting to compare our mechanism with that of random matrices i.e. the Gaussian unitary, orthogonal and symplectic ensembles etc, which can also produce evenly spaced eigenvalues within circular regions in the complex plane, akin to electrons in a quantum Hall fluid~\cite{frohlich1991large,PhysRevB.52.14137,PhysRevLett.75.697,susskind2001quantum,PhysRevLett.95.176402,Bergman_2001,jokela2011magneto,PhysRevB.90.115139,PhysRevB.99.035427}. What is markedly different is the extent of non-locality required: In our setup [Fig.\ref{fig:2}(a)], random inter-segment couplings extend across at most $M$ sites, and in the thermodynamic limit of large $N$, the entire chain can still be considered as a long PBC chain with $N$ nearest-neighbor coupled unit cells, each having a fixed $L\geq M$ number of components. However, in classical random matrix ensembles, the random elements represent all possible couplings, which in this context can be as far as $NL/2$ number of sites. Furthermore, our approach can easily be generalized to fill up arbitrarily-shaped regions by choosing the base Hamiltonian $H_\alpha$ with a similarly shaped PBC loop. 

From a more general viewpoint, our constructive approach offers an avenue for stochastically ``augmenting'' the dimensionality of a 1D system, such that it possess characteristics normally associated with 2D systems, such as 2D density of states. The probability distributions associated with the random couplings provide additional degrees of freedom that may ultimately emulate extra dimensions. Finally, we mention that our models can readily be physically implemented in media tha admit long-ranged couplings, such as classical electrical circuits~\cite{circuitPhysRevLett.114.173902,circuitSTP,circuitLee2018,circuitimhof2018topolectrical,circuitPhysRevB.99.161114,circuitPhysRevLett.122.247702,circuitPhysRevB.100.081401,circuitPhysRevB.99.020302, circuitPhysRevB.100.184202,experimenthelbig2020,lee2020imaging,circuitPhysRevLett.124.046401,circuithofmann2020reciprocal,lenggenhager2021electric,circuitliu2021non,circuitrafi2021non,circuitzou2021observation,zhang2021observation,circuitPhysRevResearch.3.023056,circuitPhysRevLett.126.215302,circuitliu202001,RevModPhys.93.015005}
 and quantum computers~\cite{garcia2020ibm,ippoliti2021many,choo2018measurement,smith2019simulating,helsen2019spectral,randall2021many,frey2021simulating,stenger2021simulating,koh2021stabilizing,experimentwu2019,PhysRevX.10.021054,PhysRevX.8.011032,madjarov2020high},
 and with some adaptation even Rydberg atom lattices with long-ranged interactions~\cite{PhysRevX.10.021054,PhysRevX.8.011032,madjarov2020high,PhysRevLett.104.173001,anderson2011trapping,PhysRevA.86.053618,bluvstein2021controlling,keesling2019quantum,omran2019generation,PhysRevLett.124.103601,liu2020localization,samajdar2021quantum,PhysRevResearch.3.043059}. Since the non-local couplings only need to be randomly distributed according to certain loosely defined distributions, our approach is intrinsically tolerant to significant levels of noise.

\bibliography{references_disorder}

\appendix
\onecolumngrid
\section{Detailed analysis of sensitivity to boundary conditions}\label{apppendix1}
We study how the simplest illustrative non-Hermitian model Eq.\eqref{eq1-2}
  \begin{equation}\label{eq1-2app}
\begin{split}
H=\sum^{L-1}_{n=1} t |n\rangle\langle n+1|+   t'|n+1\rangle\langle n|+\mu |L \rangle\langle 1|+ \mu' |1\rangle\langle L|&\  ,
\end{split}
\end{equation} 
is profoundly affected by its boundary conditions. 

Let us consider its bulk and boundary conditions, with $|\psi\rangle=\sum_n\psi_n|n\rangle$ and $H|\psi\rangle=E|\psi\rangle$:
\begin{equation}\label{eqs1-1}
\begin{split}
\quad \quad t\psi_{n+1}-E\psi_n+t'\psi_{n-1}&=0\ ,\qquad n=2,3,...,L-1\ ,\\ 
 t\psi_{2}-E\psi_{1}+\mu'\psi_{L}&=0\ ,\\
 \mu \psi_{1}-E\psi_{L}+t'\psi_{L-1}&=0\ .
\end{split}
\end{equation} 
We let $k\rightarrow k+i\kappa$, such that $E$ takes the form $E=t\exp(-\kappa+ik)+t'\exp(\kappa-ik)$ in the complex Brillouin zone. The bulk and boundary information from Eq.\eqref{eqs1-1} give rise to the following two recurrence relations
\begin{equation}\label{eqs1-2}
\begin{split}
 Q_1\begin{pmatrix}\psi_{L}\\
\psi_{L-1}
\end{pmatrix}=S_1 \begin{pmatrix}\psi_{2}\\
\psi_{1}
\end{pmatrix}\ , \qquad Q_2\begin{pmatrix}\psi_{L}\\
\psi_{L-1}
\end{pmatrix}=S_2 \begin{pmatrix}\psi_{2}\\
\psi_{1}
\end{pmatrix}\ ,\\
\end{split}
\end{equation} 
where 
\begin{equation*}
\begin{split}
Q_1=\begin{pmatrix}1&-z\\1&-z_1\end{pmatrix}\ ,\quad S_1=\begin{pmatrix}z_1^{N-2}&-zz_1^{N-2}\\z^{N-2}&-z_1z^{N-2}\end{pmatrix}\ , \quad Q_2=\begin{pmatrix}\mu'&0\\tz+t'/z&-t'\end{pmatrix}\ ,\quad S_2=\begin{pmatrix}-t&tz+t'/z\\0&\mu\end{pmatrix}\ ,
\end{split}
\end{equation*} 
with $z=\exp(-\kappa+ik),z_1=t'/t\times \exp(\kappa-ik)$.
That is, $\det(Q^{-1}_1S_1-Q^{-1}_2S_2)=0$, which yields the characteristic equation
\begin{equation}\label{eqs1-3}
\begin{split}
\left(tt'\left(\frac{t'}{t}\right)^{L}\frac{1}{z^L}- t' \mu\right)\left(t\mu' z^{L}-tt'\right)&=z^2\left(t\mu' \left(\frac{t'}{t}\right)^{L}\frac{1}{z^L}-tt'  \right)\left(t^2z^L-t \mu \right)\ .
\end{split}
\end{equation} 
In order to get the form of $\kappa$, we consider several cases.

First, considering the system with PBC ($\mu=t,\mu'=t'$), the characteristic equation Eq.\eqref{eqs1-3} takes the form 
\begin{equation}
\begin{split}
\left(tt'\left(\frac{t'}{t}\right)^{L}\frac{1}{z^L}- tt'\right)\left( z^{L}-1\right)&=t/t' z^2t/t'\left(tt'\left(\frac{t'}{t}\right)^{L}\frac{1}{z^L}- tt'\right)\left(z^L-1\right)\ ,
\end{split}
\end{equation} 
with a common factor yielding the solution $z^L=1$. Hence $\kappa=0$, $k=2\pi n/L,n=1,2,..,L$, as expected.

\indent And for the system with OBC ($\mu=0,\mu'=0$), the characteristic equation Eq.\eqref{eqs1-3} is
\begin{equation}
\begin{split}
 z^{2L+2}=\left(\frac{t'}{t}\right)^{L+1}\ ,\qquad \ z=\sqrt{t'/t}\times\text{e}^{2i\pi n/(L+1)}\ , \quad n=1,2,...,L\ ,
\end{split}
\end{equation} 
with $ z=\sqrt{t'/t}\times\text{e}^{2i\pi n/(L+1)}$, $\kappa=1/2\log(t/t')$, $k=2\pi n/(L+1), n=1,2,...,L$. The results of PBC and OBC are consistent with well-known results.\\
\indent We next consider scenarios with partially porous boundaries, i.e. with $\mu\neq0,\mu'\neq 0$ boundary hoppings.  
\begin{equation}
\begin{split}
&t'<t: z^L=\frac{\mu}{t}\frac{tt'-t^2z^2}{\mu\mu'-t^2z^2}=\frac{t'}{\mu'}\frac{tt'\mu\mu'-t^2\mu\mu'z^2}{tt'\mu\mu'-t^3t'z^2}\Rightarrow \lim\limits_{\mu\mu' < tt'}z=\sqrt[L]{\frac{\mu}{t}}\text{e}^{2i\pi n/L}, \quad \lim\limits_{\mu\mu'> tt'}z=\sqrt[L]{\frac{t'}{\mu'}}\text{e}^{2i\pi n/L}\ ,\\
&t'>t: z^L=\frac{\mu}{t}\frac{\mu'z^2-t'^2/\mu}{\mu'z^2-t'\mu'/t}=\frac{t'}{\mu'}\frac{1-\mu\mu'z^2/t'^2}{1-t z^2/t'}\Rightarrow \lim\limits_{\mu\mu'< tt'}z=\sqrt[L]{\frac{t'}{\mu'}}\text{e}^{2i\pi n/L}, \quad \lim\limits_{\mu\mu'> tt'}z=\sqrt[L]{\frac{\mu}{t}}\text{e}^{2i\pi n/L}\ .
\end{split}
\end{equation} 
the limits evaluated via the approximation $\lim\limits_{L\gg1}\sqrt[L]{x}\approx 1$ which holds for $x\geq1$ and $n=1,2,...,L$. Explicitly, 
\begin{equation}
\begin{split}
&t'<t: z=\left(\frac{\mu}{t}\right)^{1/L}\left(\frac{tt'-t^2z^2}{\mu\mu'-t^2z^2}\right)^{1/L}
\to \lim\limits_{\mu\mu' < tt'} \left(\frac{\mu}{t}\right)^{1/L}\text{e}^{2i\pi n/L}\\
& \quad \text{with}\quad \left(\frac{tt'-t^2z^2}{\mu\mu'-t^2z^2}\right)>1 \quad \text{and }\quad  \lim\limits_{L\gg 1}\left(\frac{tt'-t^2z^2}{\mu\mu'-t^2z^2}\right)^{1/L}\to 1.
\end{split}
\end{equation} 
The situation where either $\mu$ or $\mu'$ vanishes is more subtle, but actually quite similar. For $\mu\neq0,\mu'= 0$, we have
\begin{equation}
\begin{split}
& t'<t:\quad   z^{L}=\frac{\mu}{t}\frac{ z^2-t'/t}{z^2}\ , \quad  z=\sqrt[L]{\frac{\mu}{t}}\times\text{e}^{2i\pi n/L}\ , \  n=1,2,...,L\ ,
 \\
& t'>t:\quad  z^{2L+2}\approx\left(\frac{t'}{t}\right)^{L+1}\ ,  \quad z=\sqrt{\frac{t'}{t}}\times \text{e}^{2i\pi n/(L+1)}\ , \  n=1,2,...,L\ .
\end{split}
\end{equation} 
And for the case with $\mu=0,\mu'\neq 0$, we have 
\begin{equation}
\begin{split}
 &t'<t: \  z=\sqrt{\frac{t'}{t}}\times\text{e}^{2i\pi n/(L+1)}\ ,  \quad n=1,2,...,L\ ,\\
& t'>t:\   z=\sqrt[L]{\frac{t'}{\mu'}}\times\text{e}^{2i\pi n/L}\  , \qquad\    n=1,2,...,L \ .
\end{split}
\end{equation} 
Combining the above results, we obtain Table.\ref{tab1} in the main text.

\begin{figure}[htp]
    \begin{centering}
    \includegraphics[width=0.9\linewidth]{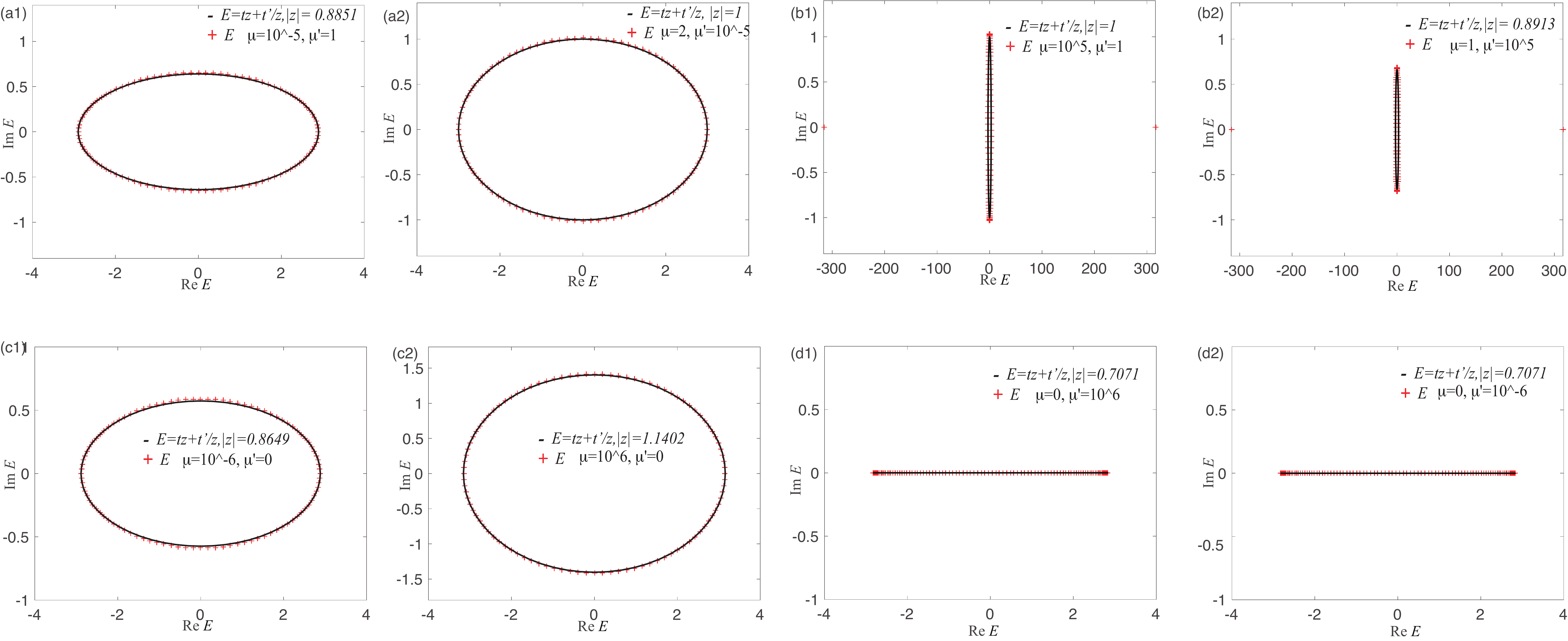}
    \par\end{centering}
    \protect\caption{\label{fig:s11} Near-perfect agreement of analytically approximated spectra (black) with numerical spectra (red) for the 4 cases discussed above: $\mu\mu' \ll tt'$(a1,a2);\ $\mu\mu' \gg tt'$(b1,b2);\ $\mu\neq 0,\mu'=0$(c1,c2);\ $\mu=0,\mu'\neq 0$(d1,d2). 
    The red crosses represent numerical results of  Hamiltonian Eq.\eqref{eq1-2app}, and the black points show the results (analytical approximate solution ) from TABLE.\ref{tab1}.  Other parameters are $t=2,\ t'=1,\ N=100$. Since $t>t'$, an exponentially small part of the state can feel $\mu'=0$ ($\mu=0$), and the system still behaves like it is under PBCs(OBCs), as in (c) and (d). Qualitatively similar conclusions apply to (a) and (b).  }
\end{figure}

 By comparing with numerically obtained spectra, Fig.\ref{fig:s11} verifies the correctness of the approximate analytical results obtained above and given in TABLE.\ref{tab1}.

\begin{figure}[htp]
    \begin{centering}
    \includegraphics[width=.8\linewidth]{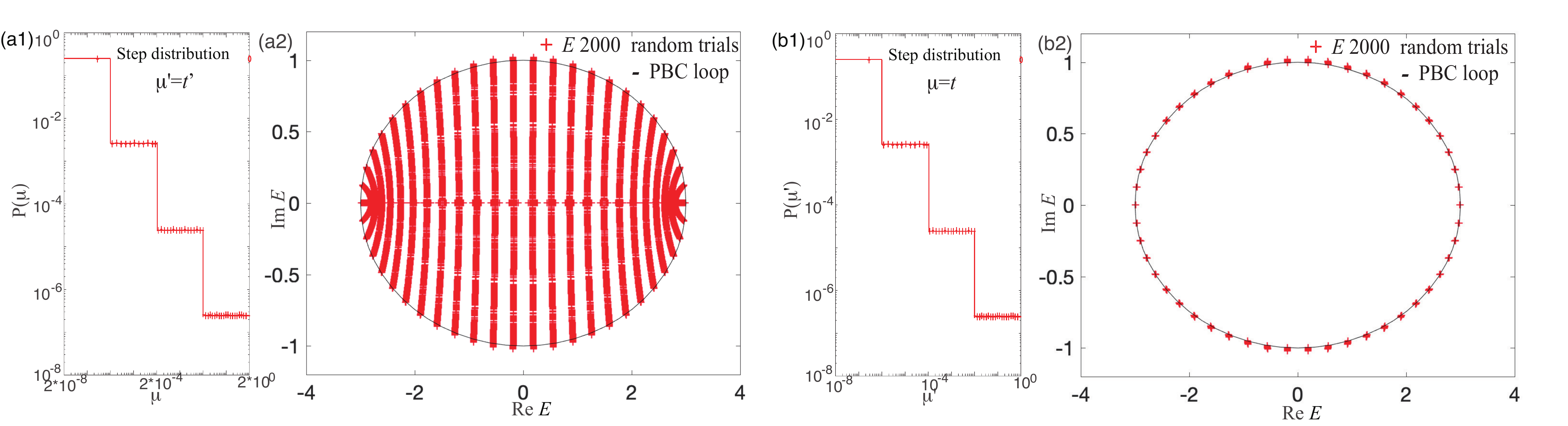}
    \par\end{centering}
    \protect\caption{\label{fig:s12} Combined energy spectrum (a2,b2) of  Hamiltonian Eq.\eqref{eq1-2app} with 2000 random trials of $\mu,\mu'$ picked from distributions $P$ as given in (a1,b1)  When $t'<t$ and $\mu\mu’ < tt’$, neither $E$ nor $z$ has anything to do with  $\mu’$.
     Other parameters are $t=2,t'=1,N=50$. 
		}

\end{figure}

Now, if the boundary hoppings $\mu,\mu'$ were to be random numbers from $(0,t)$ and $(0,t')$, the PBC energy spectrum would be ellipses of all different aspect ratios, and after multiple random trials, the  combined energy spectrum would fill up the PBC loop, as shown in Fig.\ref{fig:s12}(a2). Note that this filling is sensitive to the direction of skin mode accumulation, so if $t>t'$, only the case in Fig.\ref{fig:s12}(a2) and not that of Fig.\ref{fig:s12}(b2) will occupy the interior of the PBC loop. To maintain an approximately uniform filling density, we have concocted a step-like distribution given by
\begin{equation}
	P(x)=\left\{
	\begin{aligned}
	&10^{-6}/4\ ,\qquad x\in(10^{-2},1]\\
&10^{-4}/4\ ,\qquad x\in(10^{-4},10^{-2}]\\
&10^{-2}/4\ ,\qquad x\in(10^{-6},10^{-4}]\\
&1/4\ ,\qquad\quad\  \ x\in(0,10^{-6}]
	\end{aligned}
	\right.
	\end{equation}
	where $x$ represents either $\mu$ or $\mu'$.

\section{ Effect of relative strengths of random couplings}

We now study the effect of tuning the overall strength of the random couplings in $H_b$ by considering the parametrization
\begin{equation} H=\sum_\alpha^N H_\alpha + \lambda H_b\ .
\label{rescaled}
\end{equation}
We recover the results of the main text when $\lambda=1$, and that of the clean non-Hermitian chain when $\lambda=0$. In the latter limit, $\kappa$ must correlate poorly with the IPR, since the system is essentially that of the OBC clean system, with only a single inverse skin depth. Note that with $\lambda$, the best filling $\sigma$ amplitude is also rescaled by a factor of $\lambda$. 

\begin{figure}[h!]
    \begin{centering}
    \includegraphics[width=0.65\linewidth]{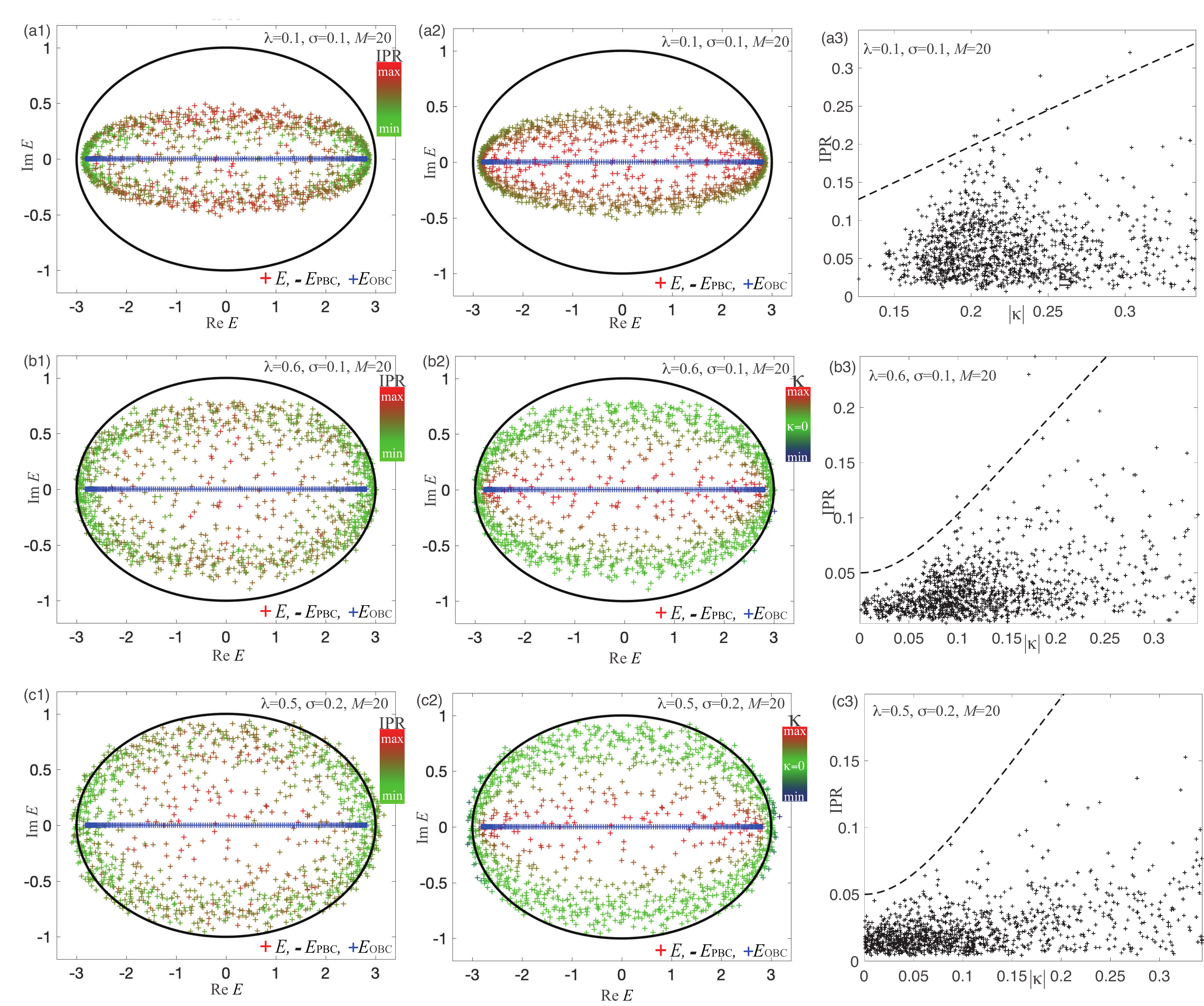}
    \par\end{centering}
    \protect\caption{\label{fig:51} IPR vs. $\kappa$ of our model under the rescaling $\lambda$ of the random couplings. 
     Eigenenergies are colored by the IPR Eq.\eqref{eq3-1} (a1-c1) or $\kappa$ Eq.\eqref{eq3-2} (a2-c2). (a3-c3)  shows the correlation between IPR and $\kappa$. The correlation, although imperfect, is best at larger disorder strengths such as $\lambda\approx 0.6$. 
}
\end{figure}

From Fig.\ref{fig:51}, the correlation is poor for small $\lambda$, as expected, since the system is not too different from an OBC system with small amounts of disorder. This is evident in the ``flattening'' of the spectrum in the complex energy plane. The correlation as well as the filling improves as $\lambda$ increases.

\begin{figure}[h!]
    \begin{centering}
    \includegraphics[width=0.65\linewidth]{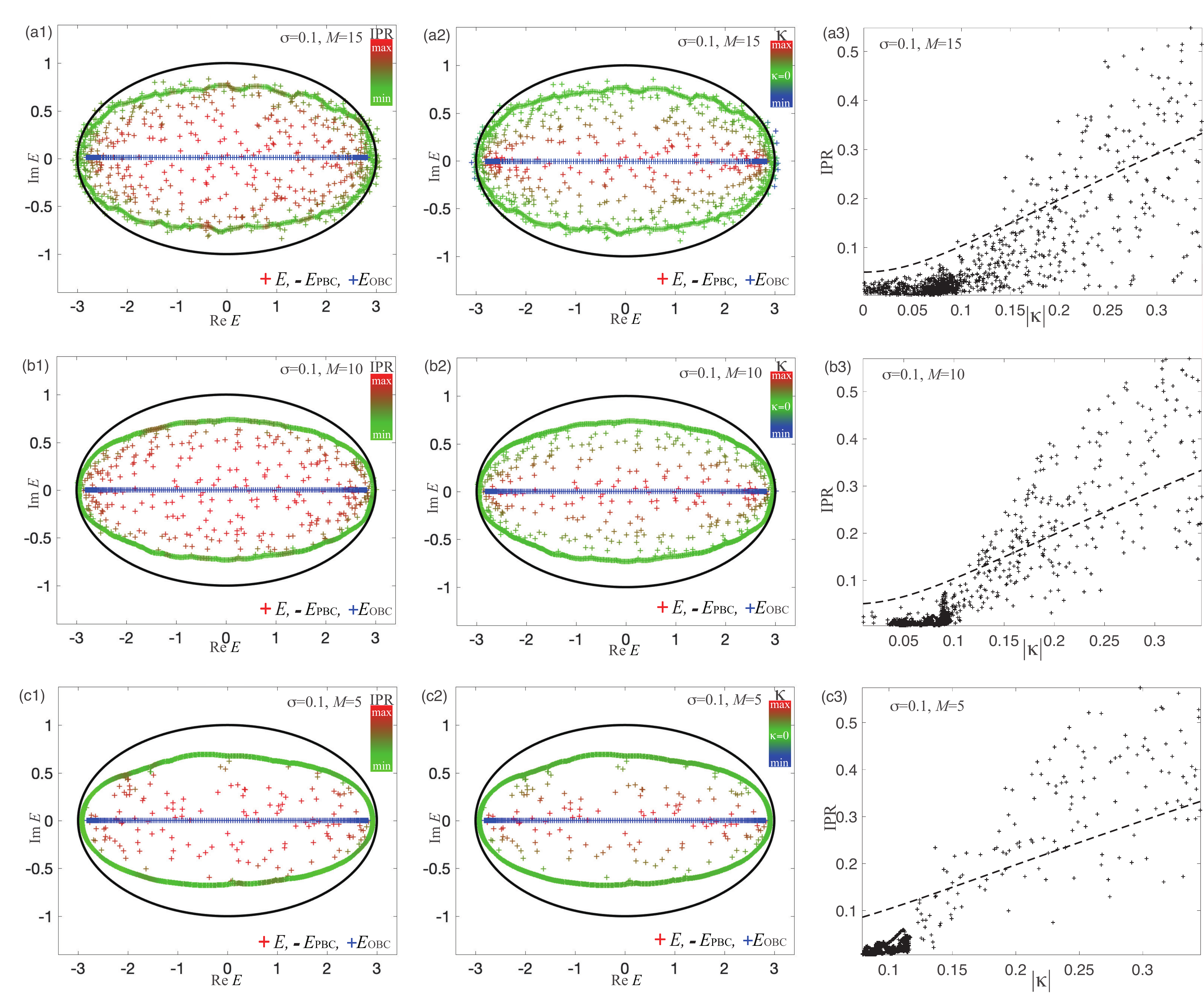}
    \par\end{centering}
    \protect\caption{\label{fig:61} IPR vs. $\kappa$ of our model ($\lambda=1$) for different maximal random coupling distances $M$. Eigenenergies are colored by the IPR Eq.\eqref{eq3-1} (a1-c1) or $\kappa$ Eq.\eqref{eq3-2} (a2-c2). (a3-c3)  shows the correlation between IPR and $\kappa$. The correlation, is best at smaller disorder coupling distances $M$.}
\end{figure}

In general, it was also found from Fig.\ref{fig:61} that the correlation improves as $M$ decreases. This is not surprising, since smaller $M$ implies fewer random couplings, thereby increasing the reliability of $\kappa$ from the base Hamiltonian $H_\alpha$ as a measure of the locality for the entire Hamiltonian $H$. That said, it is still with maximal $M=L$ that we obtain the best fillings.  

\end{document}